\documentclass[fleqn,usenatbib]{mnras}

\usepackage{newtxtext,newtxmath}
\usepackage[T1]{fontenc}
\usepackage{graphicx}
\usepackage{amsmath,mathtools}
\usepackage{bm}
\usepackage{booktabs}
\usepackage{tikz}
\usepackage{url}
\hypersetup{hidelinks}
\usetikzlibrary{decorations.pathmorphing,decorations.markings,calc,patterns,arrows.meta,positioning}
\graphicspath{{figures/}}
\newcommand{\pp}{\partial}
\newcommand{\diag}{\mathrm{diag}}

\newcommand{\norm}[1]{\left\lVert #1\right\rVert}
\newcommand{\R}{\mathbb{R}}

\title[Resolving power of effective rheologies]{Resolving power of effective rheologies of planetary interiors with an application to the Moon}

\author[Y. Gevorgyan]{
Yeva Gevorgyan$^{1}$\thanks{E-mail: yeva.gevorgyan@kaust.edu.sa}\\
$^{1}$King Abdullah University of Science and Technology (KAUST), Thuwal 23955-6900, Saudi Arabia
}

\date{Accepted XXX. Received YYY; in original form ZZZ}
\pubyear{\the\year{}}

\begin{document}
\label{firstpage}
\pagerange{\pageref{firstpage}--\pageref{lastpage}}
\maketitle

\begin{abstract}
Layered viscoelastic interiors control the frequency dependence of planetary
tidal dissipation, but repeated layered Love-number calculations are expensive
in long-term orbital and spin integrations. Equivalent homogeneous rheologies
provide a cheaper representation of the same response, but their effective
parameters do not by themselves identify the interior layers they represent. We
develop a local interpretation of such parameters by differentiating the
layered-to-homogeneous construction about a fixed-geometry baseline, perturbing
only selected layer rigidities and viscosities. We use the pole-residue
representation of the degree-two Love number to construct a
physical-to-effective transfer matrix. We then use its singular values,
resolution matrix, and mode-dominance matrix to determine which physical perturbations are visible
in the Love-number response and how they appear in effective-rheology
coordinates. Applied to a five-layer
lunar model, the full degree-two response is sensitive to seven of eight
solid-layer rheological parameters. The single near-null direction is the
inner-core rigidity, consistent with shielding by the fluid outer core and with
the weak dependence of Love numbers on core elastic structure. Over the
LLR frequency interval, fewer physical directions are visible. To first
order, the elastic spring, isolated dashpot, and two Voigt elements reproduce the nine-pole
response over this interval. Finite
perturbations confirm the tangent prediction to within a few percent, with the
validity radius limited by a nearly degenerate pair of long-timescale modes.
\end{abstract}


\section{Introduction}
\label{sec-intro}

Tidal dissipation is a primary driver of the long-term orbital and rotational evolution of planets and satellites, from the Solar System to close-in exoplanets \citep{Bagheri2022,Boue2019,Correia2014}. Its magnitude and variation with frequency depend on the deformable body's internal structure and rheology. Real interiors are radially layered, so a layered model is the natural forward description for tidal deformation and dissipation \citep{Sabadini2016,Henning2014}. The drawback of layered models is computational, since each Love-number evaluation requires a full deformation solve \citep{Tobie2005}, repeated many times in a secular (long-term, orbit-averaged) or time-domain orbital integration. These integrations are costly, so dynamical studies often resort to highly simplified homogeneous models---a single Maxwell element or a constant quality factor---that are cheap to evaluate but cannot reproduce the frequency-dependent dissipation of a stratified interior, missing features such as the additional dissipation peaks contributed by individual layers \citep{EfroimskyLainey2007,Bolmont2020,Gevorgyan2021}.

This computational bottleneck can be reduced by replacing the layered body with a homogeneous body with a complex rheology. The interior is not physically homogeneous. Instead, an effective homogeneous model reproduces the same frequency-dependent Love number. Layered bodies with finitely many Maxwell layers and a simple relaxation spectrum can be represented by equivalent generalized Maxwell or generalized Voigt models \citep{RagazzoRuiz2017,Gevorgyan2023,GevorgyanPackage2026}. We use finite-mode approximations to extend this representation to continuous-spectrum rheologies such as Andrade and Sundberg--Cooper \citep{Gevorgyan2020,GevorgyanRagazzo2026}.

For each interior, we construct an effective rheology whose parameters reproduce the Love number but carry no physical interpretation on their own. An effective viscosity or rigidity may reflect a single interior layer, several layers, or no layer in particular, and they are not simply layer averages---averaging the layer rigidity and viscosity over- or under-estimates the dissipation depending on structure, and reproducing the layered response requires fitted values instead \citep{Bolmont2020,Henning2014}.

We differentiate the pointwise reduction from a layered Maxwell body to its effective homogeneous generalized Voigt model about a chosen baseline to determine how the effective parameters change under perturbations of the physical layer rheologies. We use the resulting transfer matrix to measure how strongly each effective parameter is tied to a physical layer and which interior perturbations strongly or weakly affect the Love-number response. The geometry and density structure are held fixed and only the layer rigidities and viscosities are perturbed, which isolates the rheological differential map and matches applications where a reference geometry is chosen first and the uncertain quantities are the layer rheologies. We use the same differential to propagate this uncertainty to the effective parameters used downstream.

Implicit differentiation of simple roots, eigenvalue sensitivity, and resolution diagnostics from linear inverse theory are classical tools. We combine them with the homogeneous-rheology inversion and express the resulting derivative in effective-rheology coordinates rather than directly in observable space. This distinguishes the present construction from standard sensitivity kernels of tidal response to interior structure, including direct Love-number partial derivatives \citep{OkuboSaito1983,Tobie2005,LauAlAttar2021} and relaxation-mode kernels in glacial isostatic adjustment \citep{Peltier1976,TrompMitrovica1999,AlAttarTromp2014}.

In Section~\ref{sec-background}, we recall the pole--residue representation of
the layered Love number and the algebraic generalized Voigt inversion. We
derive the layered-model spectral sensitivities in
Section~\ref{sec-layered-sensitivity} and differentiate the homogeneous
inversion to obtain the local transfer matrix in
Section~\ref{sec-effective-map}. We give a minimal two-layer example in
Section~\ref{sec-two-layer}, apply the method to the Moon in
Section~\ref{sec-moon}, and discuss interpretation, limitations, and the
additional requirements for observational inference in
Section~\ref{sec-discussion}.

\section{The pointwise reduction and the local map}
\label{sec-background}

The reduction consists of two stages
\citep{Gevorgyan2023,GevorgyanPackage2026,GevorgyanHomogeneousModelCode2026}.
In the first, the pole--residue spectral state $u$ is obtained from the physical
layer rheologies $v$ through the Love number $k_2(s)$
(Section~\ref{subsec-layered-response}). The effective parameters $w$ are then
recovered from $u$ by the algebraic generalized Voigt inversion $\mathcal I$
(Section~\ref{subsec-voigt}). The local transfer matrix is the derivative of
the composed map $v\mapsto u(v)\mapsto w(v)$
(Section~\ref{subsec-local-question}).

\subsection{Layered tidal response}
\label{subsec-layered-response}

Consider a spherically symmetric incompressible body made of finitely many homogeneous layers, with fixed radii and densities. Solid layers are modeled with Maxwell rheology. For liquid layers, the same deformation equations are evaluated in the zero-shear limit \citep{JaraOrue2011}. The degree-two tidal Love number is obtained from the standard propagator-matrix solution of the linear deformation equations \citep{Sabadini2016}. Under the usual simple-spectrum assumptions, this gives a rational function of the Laplace variable with the partial-fraction expansion \citep[Eq.~26]{Gevorgyan2023}
\begin{equation}\label{eq-prony}
k_2(s)=k_E+\sum_{i=1}^{n+1}\frac{r_i}{s-s_i},
\end{equation}
where $0>s_1>s_2>\cdots>s_{n+1}$ are the relaxation poles and $r_i>0$ are the corresponding residues. The elastic and fluid Love numbers are the high- and zero-frequency limits,
\begin{equation}\label{eq-kE-kf-limit}
k_E=\lim_{|s|\to\infty} k_2(s),
\quad
k_f=k_2(0)=k_E-\sum_{i=1}^{n+1}\frac{r_i}{s_i}.
\end{equation}
The imaginary part of $k_2(i\omega)$ gives the dissipative part of the tidal response at forcing frequency $\omega$.

\subsection{Equivalent generalized Voigt model}
\label{subsec-voigt}

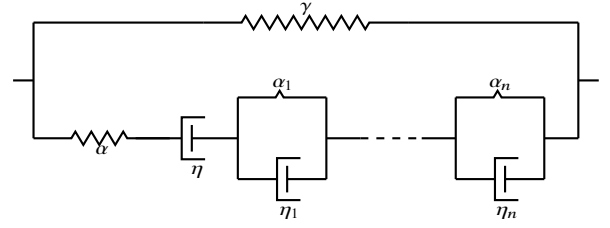
\begin{figure}
\begin{center}
\begin{tikzpicture}[scale=0.9, transform shape]
\tikzstyle{spring}=[thick, decorate, decoration={zigzag, pre length=0.5cm, post length=0.5cm, segment length=6}]
\tikzstyle{damper}=[thick, decoration={markings,
  mark connection node=dmp,
  mark=at position 0.5 with
  {
    \node (dmp) [thick, inner sep=0pt, transform shape, rotate=-90, minimum width=15pt, minimum height=3pt, draw=none] {};
    \draw [thick] ($(dmp.north east)+(5pt,0)$) -- (dmp.south east) -- (dmp.south west) -- ($(dmp.north west)+(5pt,0)$);
    \draw [thick] ($(dmp.north)+(0,-5pt)$) -- ($(dmp.north)+(0,5pt)$);
  }
}, decorate]
\tikzstyle{ground}=[fill,pattern=north east lines,draw=none,minimum width=0.75cm,minimum height=0.3cm]

            \draw [thick] (0,0.8) --  (0,2.5);

           \draw [spring] (0,0.8) -- node[below] {$\alpha$} (2,0.8);
            \draw [damper] (1.5,0.8) -- (3,0.8);
            \node at (2.4,0.3) {$\eta$};

            \draw [thick] (3,0.2) --  (3,1.4);

           \draw [spring] (3,1.4) -- node[above] {$\alpha_1$} (4.3,1.4);
            \draw [damper] (3,0.2) -- (4.3,0.2);
            \node at (3.75,-0.3) {$\eta_{1}$};

            \draw [thick] (4.3,0.2) --  (4.3,1.4);

            \draw [thick] (4.3,0.8) --  (4.8,0.8);
            \draw [thick,dashed] (4.8,0.8) -- (5.7,0.8);
            \draw [thick] (5.7,0.8) -- (6.2,0.8);

            \draw [thick] (6.2,0.2) --  (6.2,1.4);

           \draw [spring] (6.2,1.4) -- node[above] {$\alpha_n$} (7.5,1.4);
            \draw [damper] (6.2,0.2) -- (7.5,0.2);
            \node at (6.95,-0.3) {$\eta_{n}$};

            \draw [thick] (7.5,0.2) --  (7.5,1.4);

            \draw [thick] (7.5,0.8) --  (8,0.8);

            \draw [thick] (8,0.8) --  (8,2.5);

            \draw [thick] (0,2.5) --  (2.5,2.5);
           \draw [spring] (2.5,2.5) -- node[above] {$\gamma$} (5.5,2.5);
            \draw [thick] (5.5,2.5) --  (8,2.5);

            \draw [thick] (-0.3,1.65) --  (0,1.65);
            \draw [thick] (8,1.65) --  (8.3,1.65);

\end{tikzpicture}
\end{center}
\caption{Schematic representation of the generalized Voigt model used as the equivalent homogeneous body.}
\label{fig:voigt-network}
\end{figure}

The equivalent homogeneous body has a generalized Voigt rheology, following the layered-to-homogeneous construction of \citep{RagazzoRuiz2017,Gevorgyan2023,GevorgyanPackage2026}. It consists of an elastic spring $\alpha$, an isolated dashpot $\eta$, and $n$ Kelvin--Voigt elements (spring $\alpha_i$, dashpot $\eta_i$) in series, all in parallel with a gravitational spring $\gamma$ (Figure~\ref{fig:voigt-network}). The number of these elements is fixed by the pole structure of the layered response, so the resulting rheological model has the same $n+1$ relaxation poles as \eqref{eq-prony}.

The self-gravitating body compliance $C(s)$ satisfies
\begin{equation}\label{eq-k2-compliance}
k_2(s)=\Theta\,C(s), \quad
C(s)=\frac{1}{\alpha+\gamma}+C_v(s),
\end{equation}
where $\Theta=3I_0G/R^5$ is fixed by the body's geometry and density, $1/(\alpha+\gamma)$ is the instantaneous compliance, and $C_v(s)$ is the frequency-dependent viscoelastic part, with $C_v(s)\to0$ as $|s|\to\infty$ \citep{Ragazzo2020}.

The gravitational and elastic rigidities are determined by the zero- and high-frequency limits of $C(s)$ \citep{RagazzoRuiz2017,Ragazzo2020},
\begin{equation}\label{eq-gamma-alpha}
    \gamma=\frac{\Theta}{k_f},
    \quad
    \alpha=\frac{\Theta}{k_E}-\gamma.
\end{equation}
Since $\Theta$ and $k_f$ are fixed, $\gamma$ is fixed as well.

To extract the remaining viscoelastic parameters $\eta$ and $\{\alpha_i,\eta_i\}_{i=1}^{n}$, we use the complex compliance of the generalized Voigt rheology,
\begin{equation}\label{eq-Jhat}
    \widehat J(s)=\frac{1}{\alpha}+\frac{1}{\eta s}+\sum_{i=1}^{n}\frac{1}{\alpha_i+s\eta_i}.
\end{equation}

The generalized Voigt rheology is connected in parallel with the
gravitational spring. The two compliances are therefore related by
\begin{equation}\label{eq-C-Jhat-relation}
    C(s)=\frac{1}{\gamma+1/\widehat J(s)}.
\end{equation}

With $k_2=N/P$ from \eqref{eq-prony}, where $P$ and $N$ have degree $n+1$ and $P$ is monic, the parallel-spring identity gives
\begin{equation}\label{eq-Jhat-rational}
    \widehat J(s)=\frac{k_2(s)}{\Theta-\gamma\,k_2(s)}=\frac{N(s)}{\Theta\,P(s)-\gamma\,N(s)}.
\end{equation}

With $\gamma=\Theta/k_f$, the denominator in \eqref{eq-Jhat-rational} has a simple zero at $s=0$. We write
\begin{equation}\label{eq-Q-def}
\Theta\,P(s)-\gamma\,N(s)=s\,Q(s),
\end{equation}
where $Q$ has degree $n$ and negative simple roots $q_1,\ldots,q_n$. The polynomial $Q$ is used with this normalization, not as a monic polynomial. Its leading coefficient is $\Theta-\gamma k_E=k_E\alpha$. Residue matching in \eqref{eq-Jhat-rational} gives
\begin{equation}\label{eq-voigt-inversion}
    \eta=\frac{Q(0)}{N(0)}, \quad
    \eta_i=\frac{q_i\,Q'(q_i)}{N(q_i)},\quad \alpha_i=-q_i\,\eta_i.
\end{equation}

Equations \eqref{eq-gamma-alpha} and \eqref{eq-voigt-inversion} define the pointwise inversion map $\mathcal I$ from the spectral data $(k_E,s_i,r_i)$ to the generalized Voigt parameters. We differentiate this map in Section~\ref{sec-effective-map}.

\subsection{The transfer matrix}
\label{subsec-local-question}

Let $v\in\R^p$ be the physical parameter vector formed by the viscosities and rigidities of the solid Maxwell layers. With $\ell$ solid layers, we take
\begin{equation}\label{eq-v-general}
v=(\eta^{(1)},\mu^{(1)},\ldots,\eta^{(\ell)},\mu^{(\ell)})^T,\quad p=2\ell,
\end{equation}
where parenthesized superscripts index physical layers and distinguish the layer parameters $\mu^{(k)},\eta^{(k)}$ from the effective parameters $\alpha_i,\eta_i$. Liquid layers are part of the fixed reference model and are not perturbed.

From $v$, we compute the spectral state through the Love number $k_2(s;v)$,
\begin{equation}\label{eq-u-local}
u=(k_E,s_1,r_1,\ldots,s_{n+1},r_{n+1})^T,
\end{equation}
and then apply the inversion $\mathcal I$ to obtain
\begin{equation}\label{eq-w-local}
w=(\gamma,\alpha,\eta,\alpha_1,\eta_1,\ldots,\alpha_n,\eta_n)^T.
\end{equation}
We linearize the composition $v\mapsto u(v)\mapsto w(v)$ about a baseline model $v_0$, with induced spectral state $u_0=u(v_0)$ and effective parameters $w_0=\mathcal I(u_0)$.

We assume that the simple-spectrum assumptions of Section~\ref{subsec-layered-response} hold at $v_0$. This is the generic case for finite incompressible multilayer Maxwell bodies under the usual tidal-stability conditions, excluding density inversions between adjacent layers that can trigger positive roots \citep[Sec.~1.8]{Sabadini2016}. Related stability analyses are given by \citet{WuPeltier1982} and \citet{Tanaka2006}. By the interlacing property of \citet[Eq.~22]{Gevorgyan2023}, the roots of $Q$ are simple. With positive residues and fixed pole and element ordering, no effective element merges with another or carries zero weight. Under these assumptions, the forward spectral map $w\mapsto u$ is smooth with nonsingular Jacobian $J_w=\pp u/\pp w$, so its inverse $\mathcal I:u\mapsto w$ is a local diffeomorphism at $u_0$.

Throughout the paper, $Df(x_0)$ denotes the Fr\'{e}chet derivative of a smooth map $f$ at $x_0$, i.e., the Jacobian matrix $\partial f/\partial x(x_0)$. The derivative of the local reduction is written
\begin{equation}\label{eq-transfer-first}
\Delta w\approx T\Delta v.
\end{equation}
Since $w=\mathcal I(u)$, the chain rule gives
\begin{equation}\label{eq-transfer-chain}
T=D\mathcal I(u_0)Du(v_0).
\end{equation}
Equivalently, differentiating the local inverse relation between the generalized Voigt parameters and the spectral data gives
\begin{equation}\label{eq-transfer-Jw}
T=J_w^{-1}J_v,\quad
J_w=\frac{\pp u}{\pp w}(w_0),\quad
J_v=\frac{\pp u}{\pp v}(v_0).
\end{equation}

The algebraic inversion uses the complete coordinate sets $(k_E,s_i,r_i)\mapsto(\gamma,\alpha,\eta,\alpha_i,\eta_i)$, which have the same dimension. For fixed geometry and density, $Dk_f=0$, hence $D\gamma=0$, and one may drop the $\gamma$ row and recover $k_E$ from the pole-residue variables using the fixed value of $k_f$ in \eqref{eq-kE-kf-limit}. On the resulting $2n+2$ coordinates, $J_w$ is square and nonsingular, so the pseudoinverse $J_w^{+}$ used in the implementation coincides with $J_w^{-1}$.

\section{Spectral sensitivities of the layered model}
\label{sec-layered-sensitivity}

The spectral Jacobian $J_v=Du(v_0)$ has rows given by the Fr\'{e}chet derivatives of the spectral coordinates $k_E$, $s_i$, and $r_i$ with respect to the physical layer parameters. Standard sensitivity analyses often differentiate the Love number itself, giving direct or adjoint kernels for $Dk_2(s;v_0)$ \citep{OkuboSaito1983,LauAlAttar2021}. Here we instead differentiate the pole-residue coordinates of $k_2$, which gives the mode-resolved input to the effective-rheology inversion of Section~\ref{sec-effective-map}.

For a finite multilayer Maxwell body, the propagator-matrix construction gives a rational Love number $k_2(s,v)=N(s,v)/P(s,v)$, where $N$ and $P$ are polynomials in $s$ \citep{Sabadini2016,Gevorgyan2023}. The $n+1$ roots of $P$ are the relaxation poles. In monic form,
\begin{equation}\label{eq-polynomial}
P(s,v)=s^{n+1}+b_n(v)s^n+b_{n-1}(v)s^{n-1}+\cdots+b_1(v)s+b_0(v),
\end{equation}
so that $P(s,v_0)=\prod_{i=1}^{n+1}(s-s_i)$ at the baseline.

The physical rheological parameters enter through the complex shear moduli of the solid layers,
\begin{equation}\label{eq-maxwell-modulus}
\widehat\mu_k(s)=\frac{\mu^{(k)}\eta^{(k)}s}{\mu^{(k)}+\eta^{(k)}s},
\end{equation}
with
\begin{equation}\label{eq-maxwell-modulus-derivatives}
\frac{\pp\widehat\mu_k}{\pp\mu^{(k)}}=
\frac{(\eta^{(k)})^2s^2}{(\mu^{(k)}+\eta^{(k)}s)^2},
\quad
\frac{\pp\widehat\mu_k}{\pp\eta^{(k)}}=
\frac{(\mu^{(k)})^2s}{(\mu^{(k)}+\eta^{(k)}s)^2}.
\end{equation}
Let $F(s,\widehat\mu_1,\ldots,\widehat\mu_\ell)$ denote the characteristic determinant of the propagator-matrix solution. At fixed $s$, its derivative with respect to a physical parameter is
\begin{equation}\label{eq-chain-Pv}
    \frac{\pp F}{\pp v_j}\bigg|_{s}=\sum_k \frac{\pp F}{\pp\widehat\mu_k}\,\frac{\pp\widehat\mu_k}{\pp v_j}.
\end{equation}
The derivatives $\pp F/\pp\widehat\mu_k$ are obtained from the propagator-matrix construction, either symbolically or by finite differences at the baseline. After clearing the Maxwell denominators, removing common factors, and normalizing $P$ to be monic, this gives the derivatives $P_{v_j}$ used below. The same procedure applied to the numerator determinant gives $N_{v_j}$.

We first compute the pole sensitivities. Each pole is a simple root of the characteristic equation, equivalently $P(s_i(v),v)=0$. Implicit differentiation gives
\begin{equation}\label{eq-root-sens-general}
\frac{\pp s_i}{\pp v_j}
=
-\frac{P_{v_j}(s_i,v_0)}{P_s(s_i,v_0)}.
\end{equation}
This quotient is invariant under multiplication of $P$ by any factor that is nonzero at the pole, including factors that depend on $s$ or $v$, because $P=0$ at the pole. It may therefore be evaluated either in the monic normalization or directly from the cleared characteristic determinant, provided no pole coincides with a cleared Maxwell factor and the common-factor structure is locally unchanged.

Equivalently, \eqref{eq-root-sens-general} is the simple-eigenvalue sensitivity formula for the companion matrix of \eqref{eq-polynomial} \citep{Lancaster1964,Frank1978,Nelson1976}. If poles cluster or coincide, the scalar quotient breaks down and one must differentiate the corresponding invariant subspace instead \citep{Nelson1976}. This degenerate case is excluded by the simple-spectrum assumptions.

For a simple pole $s_i(v)$ of $k_2=N/P$, the corresponding residue is
\begin{equation}\label{eq-residue-general}
r_i(v)=\frac{N(s_i(v),v)}{P_s(s_i(v),v)}.
\end{equation}
Differentiating this expression gives
\begin{equation}\label{eq-residue-sensitivity-general}
\frac{\pp r_i}{\pp v_j}
=
\frac{N_{v_j}+N_s\pp_{v_j}s_i}{P_s}
-
\frac{N\left(P_{sv_j}+P_{ss}\pp_{v_j}s_i\right)}{P_s^2}.
\end{equation}
For the Jacobian $J_v$, this expression is then evaluated at $v=v_0$ and $s_i=s_i(v_0)$.

The elastic coordinate $k_E$ is the high-frequency limit of $k_2(s,v)$. Since $\widehat\mu_k(s)$ tends to the instantaneous rigidity $\mu^{(k)}$ as $|s|\to\infty$,
\begin{equation}\label{eq-kE-elastic}
k_E(v)=k_2^{\rm elastic}\bigl(\{\mu^{(k)}\}_{k=1}^{\ell}\bigr).
\end{equation}
Thus $k_E$ is computed from the elastic body obtained by replacing each complex shear modulus by its elastic rigidity. It depends on the rigidities but not on the viscosities, so
\begin{equation}\label{eq-kE-sens}
\frac{\pp k_E}{\pp \mu^{(k)}}(v_0)
=
\frac{\pp k_2^{\rm elastic}}{\pp \mu^{(k)}}\bigg|_{\mu^{(m)}=\mu^{(m)}_0,\; m=1,\ldots,\ell},
\quad
\frac{\pp k_E}{\pp \eta^{(k)}}(v_0)=0.
\end{equation}
The rigidity derivative on the right is evaluated symbolically or by finite differences on the elastic propagator solution at the baseline.

The fluid Love number $k_f$ is not included in $u$, but it is determined by the spectral data through \eqref{eq-kE-kf-limit}. Since only the solid rheologies are perturbed here and the geometry and density are held constant, $k_f$ does not vary. The spectral derivatives must therefore satisfy
\begin{equation}\label{eq-kf-consistency}
0=
\frac{\pp k_E}{\pp v_j}
-
\sum_{i=1}^{n+1}
\left(
\frac{1}{s_i}\frac{\pp r_i}{\pp v_j}
-
\frac{r_i}{s_i^2}\frac{\pp s_i}{\pp v_j}
\right).
\end{equation}
We use this identity to check the numerical implementation.

Collecting the derivatives gives the physical spectral Jacobian
\begin{equation}\label{eq-Jv-def}
J_v=Du(v_0)\in\R^{(2n+3)\times p}.
\end{equation}
To make numerical diagnostics independent of physical units, we use relative perturbations,
\begin{equation}\label{eq-relative-vars}
\delta v_j=\frac{\Delta v_j}{v_{0,j}},
\quad
\delta u_k=\frac{\Delta u_k}{u_{0,k}},
\end{equation}
whenever the corresponding baseline components are nonzero. The scaled Jacobian is
\begin{equation}\label{eq-scaled-Jv}
\widehat J_v=\diag(u_0)^{-1}J_v\diag(v_0).
\end{equation}
Since the poles are negative, the relative pole perturbations are signed. Equivalently, $\Delta s_i/s_i$ may be read as $\Delta\log(-s_i)$ and $\Delta r_i/r_i$ as $\Delta\log r_i$, so the scaled sensitivities are logarithmic derivatives. All singular-value and resolution diagnostics below are defined using scaled variables unless stated otherwise.

\section{Differentiating the effective homogeneous reduction}
\label{sec-effective-map}

Section~\ref{sec-layered-sensitivity} gives the physical spectral Jacobian
$J_v=Du(v_0)$. We now differentiate the algebraic inversion
$u\mapsto w=\mathcal I(u)$ and compose the two differentials to obtain the
local physical-to-effective map introduced in \eqref{eq-transfer-Jw}. Thus
\begin{equation}\label{eq-scaled-T}
    \widehat T=\diag(w_0)^{-1}\,D\mathcal I(u_0)\,J_v\,\diag(v_0),
\end{equation}
where the entries of $\widehat T$ measure relative changes of effective
parameters induced by relative changes of physical layer rheologies. The
inversion $u\mapsto w$ is algebraic, so its derivative can be computed
directly from the pole-residue representation and the generalized Voigt
inversion formulas.

\subsection{Jacobian of the inversion}
\label{subsec-inversion-jacobian}

We compute the Jacobian of the pointwise inversion at the baseline spectral
state $u_0$ by differentiating the pole-residue reconstruction and the
generalized Voigt formulas. For $u=(k_E,s_1,r_1,\ldots,s_{n+1},r_{n+1})$, write
\begin{equation}\label{eq-PN-from-u}
    P(s)=\prod_{i=1}^{n+1}(s-s_i),
    \quad
    N(s)=k_E P(s)+\sum_{i=1}^{n+1}r_i\prod_{j\neq i}(s-s_j),
\end{equation}
so that these polynomials reconstruct the rational Love number. Using the
normalization of \eqref{eq-Q-def}, we regard
\begin{equation}
    Q(s,u)=\frac{\Theta P(s)-\gamma(u)N(s)}{s},
    \quad
    \gamma(u)=\frac{\Theta}{k_f},
\end{equation}
as a polynomial in $s$ depending on the spectral coordinates $u$. The division
by $s$ is regular because the numerator vanishes at $s=0$.

Using \eqref{eq-gamma-alpha} and \eqref{eq-kE-kf-limit}, the gravitational
and elastic rows are
\begin{equation}\label{eq-Dgamma-Dalpha}
\begin{aligned}
    \frac{\pp\gamma}{\pp k_E}
    &=-\frac{\Theta}{k_f^2}, \quad
    \frac{\pp\gamma}{\pp r_i}
    =\frac{\Theta}{k_f^2s_i}, \quad
    \frac{\pp\gamma}{\pp s_i}
    =-\frac{\Theta}{k_f^2}\frac{r_i}{s_i^2},
    \\
    \frac{\pp\alpha}{\pp u_k}
    &=
    -\frac{\Theta}{k_E^2}\frac{\pp k_E}{\pp u_k}
    -
    \frac{\pp\gamma}{\pp u_k}.
\end{aligned}
\end{equation}
Here $\pp k_E/\pp u_k$ equals one when $u_k=k_E$ and zero otherwise.

The isolated dashpot row follows from $\eta=Q(0)/N(0)$:
\begin{equation}\label{eq-Deta}
    \frac{\pp\eta}{\pp u_k}
    =
    \frac{Q_{u_k}(0)N(0)-Q(0)N_{u_k}(0)}{N(0)^2}.
\end{equation}
Here $Q_{u_k}(0)$ is obtained by differentiating the polynomial $Q$ after the
division in \eqref{eq-Q-def}. Equivalently,
$Q(0)=\Theta P'(0)-\gamma N'(0)$.

The simple roots $q_i$ of $Q$ define the Kelvin--Voigt elements:
\begin{equation}
    Q(q_i(u),u)=0.
\end{equation}
Implicit differentiation gives
\begin{equation}\label{eq-q-sens}
\begin{aligned}
    \frac{\pp q_i}{\pp u_k}
    &=
    -\frac{Q_{u_k}(q_i)}{Q'(q_i)},\\
    Q_{u_k}(s)
    &=
    \frac{\Theta P_{u_k}(s)
    -\gamma_{u_k}N(s)-\gamma N_{u_k}(s)}{s},
    \quad s\neq0.
\end{aligned}
\end{equation}
Here $Q'=Q_s$, the partial derivatives $P_{u_k}$ and $N_{u_k}$ are obtained
from \eqref{eq-PN-from-u}, and $\gamma_{u_k}$ is given by
\eqref{eq-Dgamma-Dalpha}. The expression is evaluated only at the nonzero
roots $q_i$.

Using
\begin{equation}
    \eta_i=\frac{q_iQ'(q_i)}{N(q_i)},
\end{equation}
the chain rule gives
\begin{equation}\label{eq-Deta-i}
\begin{aligned}
    \frac{\pp\eta_i}{\pp u_k}
    ={}&
    \frac{
    q_i Q'_{u_k}(q_i)N(q_i)
    -q_iQ'(q_i)N_{u_k}(q_i)
    }{N(q_i)^2} \\
    &+
    \left[
    \frac{Q'(q_i)+q_iQ''(q_i)}{N(q_i)}
    -
    \frac{q_iQ'(q_i)N'(q_i)}{N(q_i)^2}
    \right]
    \frac{\pp q_i}{\pp u_k},
\end{aligned}
\end{equation}
where $Q'_{u_k}=\pp_s Q_{u_k}$. Since $\alpha_i=-q_i\eta_i$, the spring
row is
\begin{equation}\label{eq-Dalpha-i}
    \frac{\pp\alpha_i}{\pp u_k}
    =
    -\frac{\pp q_i}{\pp u_k}\eta_i
    -q_i\frac{\pp\eta_i}{\pp u_k}.
\end{equation}

Equations \eqref{eq-Dgamma-Dalpha}--\eqref{eq-Dalpha-i} assemble the rows of
$D\mathcal I(u_0)=J_w^{-1}$ using all spectral and effective coordinates.

\subsection{Composition with the physical tangent space}
\label{subsec-composition-tangent}

The Jacobian derived above is the differential of the full algebraic inversion
with respect to the spectral coordinates. The physical perturbations considered
in this paper occupy a smaller tangent space, because the geometry and density
are fixed and only the solid-layer rheologies are varied. In particular, the
fluid Love number does not change. Therefore, for every physical perturbation
$\Delta v$,
\begin{equation}\label{eq-fixed-kf-tangent}
    Dk_f(v_0)\Delta v=0.
\end{equation}
Equivalently, the spectral Jacobian $J_v$ satisfies the fixed-$k_f$ identity
\eqref{eq-kf-consistency}.

Composing the algebraic inverse differential with the physical spectral
Jacobian gives
\begin{equation}\label{eq-T-composition}
    T=D\mathcal I(u_0)J_v.
\end{equation}
In relative variables this is the scaled map \eqref{eq-scaled-T}. Although
$D\mathcal I(u_0)$ contains a nonzero $\gamma$ row as a derivative with
respect to arbitrary spectral perturbations, that row vanishes after
composition with $J_v$. Indeed,
\begin{equation}
    D\gamma(u_0)J_v\Delta v
    =
    -\frac{\Theta}{k_f^2}Dk_f(v_0)\Delta v
    =
    0.
\end{equation}
For rheological perturbations, the physical-to-effective map therefore has no
gravitational component. Fixing the geometry and density imposes $Dk_f=0$, so
we work on the corresponding spectral tangent subspace.

In the implementation, the full Jacobian
$D\mathcal I(u_0)$ is computed using the coordinates
$(k_E,s_i,r_i)\mapsto(\gamma,\alpha,\eta,\alpha_i,\eta_i)$. The
diagnostics below are evaluated after composition with $J_v$ on the fixed-$k_f$
physical tangent space. We check the implementation by verifying that the
composed $\gamma$ row is negligible, as required by
\eqref{eq-kf-consistency}, and by comparing $J_w^{-1}J_v$ with the symbolic
composition $D\mathcal I(u_0)J_v$. Here $J_w=\pp u/\pp w$ is computed from the
effective forward model.

\subsection{Reduced and projected maps}
\label{subsec-reduced-maps}

Above, we derived the differential map for the complete algebraic reduction. In
applications, the effective model may be reduced before it is used in an orbital or spin integration. We select the effective parameters at the baseline and keep them fixed during differentiation. This is the local form of the dominant-mode selection used in the pointwise reduction \citep{Gevorgyan2023,GevorgyanPackage2026}. Keeping the selection fixed avoids discontinuities when a perturbation moves an element across the dominance threshold.

Let $w_{\rm red}$ be the retained effective parameter vector. The target
against which the reduced model is matched may be the reduced spectral state,
the full spectral state, or a finite set of Love-number samples over a
frequency interval. We represent this target by a differentiable map
\begin{equation}
    \mathcal G:u\mapsto y,
\end{equation}
where the real and imaginary parts are treated as separate components when $y$
is complex. The corresponding
physical and effective tangent maps are
\begin{equation}\label{eq-target-jacobians}
    J_{v,\mathcal G}=D\mathcal G(u_0)J_v,
    \quad
    J_{w,\mathcal G}=D\mathcal G(u_0)J_{w,{\rm red}}.
\end{equation}
Here $J_{w,{\rm red}}=\pp u/\pp w_{\rm red}$ is the spectral Jacobian with
respect to the retained effective parameters, evaluated by freezing the omitted
parameters at their baseline values.

If the reduced Jacobian is square and nonsingular for the chosen target, the
reduced transfer matrix is
\begin{equation}\label{eq-T-red-square}
    T_{\rm red}=J_{w,\mathcal G}^{-1}J_{v,\mathcal G}.
\end{equation}
More generally, when the retained model has fewer effective parameters than the
target requires, we use the least-squares tangent projection
\begin{equation}\label{eq-T-pinv}
    \Delta w_{\rm red}
    \approx
    J_{w,\mathcal G}^{+}J_{v,\mathcal G}\Delta v,
\end{equation}
where $J_{w,\mathcal G}^{+}$ is the Moore--Penrose pseudoinverse
\citep{Barata2012}. The pseudoinverse gives the first-order change in the
reduced effective parameters that minimizes the target mismatch in the chosen
norm. This is a local least-squares projection for the selected reduced model.

\subsection{Diagnostics}
\label{subsec-diagnostics}

We interpret the local maps using singular values, resolution matrices,
projection residuals, and mode-dominance weights. All diagnostics use
relative physical perturbations. When the target is an effective-parameter
vector, the effective parameters are scaled relatively. When the
target is the complex Love number sampled over a frequency interval, we treat
its real and imaginary parts as separate components without rescaling them
individually. Although $k_2$ is
dimensionless, its sampled real and imaginary parts can have very different scales.

Let $A$ be any scaled tangent map whose columns correspond to the physical
rheological parameters. For the full effective reduction, $A=\widehat T$. For a
reduced effective model, $A$ is the corresponding scaled reduced map. For a
direct frequency-interval diagnostic, $A$ may instead be the scaled physical map
from layer perturbations to Love-number samples. We write the singular value
decomposition as
\begin{equation}\label{eq-svd}
    A=U\Sigma V^T,
\end{equation}
and use the right singular vectors in $V$ to form combinations of physical-layer
perturbations and rank these directions by their corresponding singular values,
in decreasing order of their first-order influence on the chosen target. When the target is an effective-parameter vector, the left
singular vectors in $U$ are combinations of effective rheological parameters.
Retaining the leading right singular directions follows the dimension-reduction
construction of \citet{ConstantineDowWang2014}, commonly called an active
subspace.

To localize the physical directions retained by the map, we use the truncated
resolution matrix
\begin{equation}\label{eq-resolution}
    R_\tau=A_\tau^{+}A=V_rV_r^T.
\end{equation}
Here $A_\tau^{+}$ is the pseudoinverse formed after discarding
all singular values with $\sigma_i\leq\tau\sigma_1$, and $V_r$ collects the
retained right singular vectors. The truncation is part of the diagnostic. If
$A$ has full column rank and no truncation is applied, then $A^+A=I$ and the
resolution matrix carries no information. We therefore choose $\tau$ inside a
clear gap in the singular spectrum. A diagonal entry of $R_\tau$ close to one
means that the corresponding physical parameter lies in the retained subspace.
A value close to zero means that the direction is nearly invisible.

Here $R_\tau$ measures model-to-model resolution: it determines which physical
perturbations can be recovered from their changes to the chosen target.
Observational resolution would additionally require a data operator and noise
model \citep{BackusGilbert1970,Tarantola2005}.

For projected reduced models, we measure how much of the physical target
sensitivity is not spanned by the retained effective tangent space. The
projection residual is
\begin{equation}\label{eq-residual}
    \varepsilon_{\mathcal G}
    =
    \frac{
    \norm{
    \widehat J_{v,\mathcal G}
    -
    \widehat J_{w,\mathcal G}
    \widehat J_{w,\mathcal G}^{+}
    \widehat J_{v,\mathcal G}
    }_F
    }{
    \norm{\widehat J_{v,\mathcal G}}_F
    }.
\end{equation}
The hats indicate that scaling is applied before the pseudoinverse is formed.
Scaling and pseudoinversion do not commute in general.
The value $\varepsilon_{\mathcal G}$ measures the fraction of first-order target
sensitivity that the retained effective model cannot reproduce.

When fitting the complex Love number over a frequency interval, we combine the
Jacobians of its real and imaginary parts in a single least-squares problem. Let
$\widehat T_{\mathcal B}$ be the resulting reduced transfer matrix. The corresponding
complex residual matrix is
\begin{equation}
    \mathcal R_{\mathcal B}
    =
    \widehat J_{v,\mathcal B}
    -\widehat J_{w,\mathcal B}\widehat T_{\mathcal B}.
\end{equation}
We define
\begin{equation}\label{eq-component-residuals}
\varepsilon_{\mathcal B}^{\operatorname{Re}}
=\frac{\|\operatorname{Re}\mathcal R_{\mathcal B}\|_F}
{\|\operatorname{Re}\widehat J_{v,\mathcal B}\|_F},
\qquad
\varepsilon_{\mathcal B}^{\operatorname{Im}}
=\frac{\|\operatorname{Im}\mathcal R_{\mathcal B}\|_F}
{\|\operatorname{Im}\widehat J_{v,\mathcal B}\|_F}.
\end{equation}
We evaluate both residuals using $\widehat T_{\mathcal B}$ from the combined
least-squares problem.

We compute the mode-dominance matrix from the relative pole sensitivities in
Section~\ref{sec-layered-sensitivity}:
\begin{equation}\label{eq-dominance}
    D_{ij}
    =
    \frac{
    \left|
    \dfrac{v_{0,j}}{s_{i,0}}
    \dfrac{\pp s_i}{\pp v_j}
    \right|
    }{
    \sum\limits_{m=1}^p
    \left|
    \dfrac{v_{0,m}}{s_{i,0}}
    \dfrac{\pp s_i}{\pp v_m}
    \right|
    }.
\end{equation}
For each mode, a row gives the relative contribution of every physical
rheological parameter to its timescale sensitivity. If the denominator
vanishes, we leave the row undefined or set it to zero in plots because none of
the selected parameters changes that pole at first order.

\section{A two-layer toy model}
\label{sec-two-layer}

We use a two-layer body as a toy model.

\subsection{The two-layer model}
\label{subsec-two-model}

Consider an incompressible body with an effectively inviscid liquid core and a Maxwell viscoelastic mantle. The body and core radii, mass, and densities are fixed, so the only rheological parameters allowed to vary are those of the mantle,
\begin{equation}\label{eq-x-two}
    v=(\eta_m,\mu_m)^T,
\end{equation}
which enter the response through the mantle complex shear modulus
\begin{equation}\label{eq-maxwell-two}
    \widehat\mu_m(s)=\frac{\mu_m\eta_m s}{\mu_m+\eta_m s}
    =\frac{\mu_m s}{s+\omega_m},
    \quad
    \omega_m=\frac{\mu_m}{\eta_m}.
\end{equation}

The characteristic determinant is a quadratic polynomial in the mantle complex shear modulus $\widehat\mu_m$. After substituting \eqref{eq-maxwell-two}, the characteristic equation can be written as
\begin{equation}\label{eq-two-char}
    A(v)\,s^2+B(v)\,s+C(v)=0.
\end{equation}
The corresponding Love number is
\begin{equation}\label{eq-two-rational}
    k_2(s)=\frac{N_2(v)\,s^2+N_1(v)\,s+N_0(v)}{A(v)\,s^2+B(v)\,s+C(v)},
\end{equation}
whose coefficients $A,B,C$ and $N_0,N_1,N_2$ depend on $\eta_m$ and $\mu_m$ through $\widehat\mu_m$ and $\omega_m=\mu_m/\eta_m$. The Love number has two relaxation poles, corresponding to a generalized Voigt model with one Voigt element.

\subsection{Spectral sensitivities}
\label{subsec-two-sens}

For $v_j\in\{\eta_m,\mu_m\}$, applying \eqref{eq-root-sens-general} to the
quadratic characteristic polynomial gives
\begin{equation}\label{eq-two-root-sens}
    \frac{\pp s_i}{\pp v_j}
    =-\frac{A_{v_j}s_i^2+B_{v_j}s_i+C_{v_j}}{2As_i+B},
    \quad
    i\in\{+,-\}.
\end{equation}
We compute the residues as
\begin{equation}\label{eq-two-residue}
    r_i=\frac{N_2s_i^2+N_1s_i+N_0}{2As_i+B}=\frac{N(s_i)}{P_s(s_i)},
\end{equation}
and their derivatives using \eqref{eq-residue-sensitivity-general}. Together with $k_E$, these derivatives determine the two-layer physical spectral Jacobian $J_v$ for the variables in \eqref{eq-x-two}.

\subsection{Two-layer transfer matrix}
\label{subsec-two-transfer}

The complete spectral and effective parameter vectors have equal dimension:
\begin{equation}
    u_{2L}=(k_E,s_+,r_+,s_-,r_-)^T,
    \quad
    w_{2L}=(\gamma,\alpha,\eta,\alpha_1,\eta_1)^T.
\end{equation}
The transfer matrix is
\begin{equation}\label{eq-two-T}
    \Delta w_{2L}\approx T_{2L}\,\Delta v,
    \quad
    T_{2L}=D\mathcal I(u_{2L,0})\,J_v.
\end{equation}
For the numerical toy model, an effectively inviscid liquid core with radius $325$~km and density $6700$~kg\,m$^{-3}$ is surrounded by a homogeneous Maxwell mantle extending to the surface at $R=1737.15$~km. We assign the mantle the upper-mantle values from Table~\ref{tab-lunar-baseline}: density $3356$~kg\,m$^{-3}$, rigidity $\mu_m=62.5$~GPa, and viscosity $\eta_m=10^{21}$~Pa\,s. The two poles correspond to relaxation times $\tau_i=-1/s_i$ of $3.0\times10^{4}$~yr and $3.4\times10^{5}$~yr. With the geometry and density fixed, $k_f$ and $\gamma$ do not vary. We evaluate the scaled transfer matrix numerically at this baseline. The $\gamma$ row is below $2\times10^{-8}$ and is omitted:
\begin{equation}\label{eq-two-T-instance}
    \widehat T_{2L}=\;
    \bordermatrix{~ & \eta_m & \mu_m\cr
      \alpha   & 1\times10^{-4} & 1.0000\cr
      \eta     & 1.0000 & -4\times10^{-8}\cr
      \alpha_1 & -3\times10^{-4} & -1.6\times10^{-3}\cr
      \eta_1   & 0.9997 & -3.1\times10^{-3}\cr}.
\end{equation}
A relative change in mantle rigidity produces the same relative change in $\alpha$. A relative change in mantle viscosity produces nearly the same relative changes in $\eta$ and $\eta_1$. All other matrix entries have magnitudes below $3.1\times10^{-3}$. The small sensitivity of $\alpha_1$ indicates that it is determined mainly by the fixed geometry and density. In this model, $\alpha_1$ measures the strength of the slow core--mantle buoyancy response, and $\eta_1$ controls its relaxation time.

\section{Five-layer model of the Moon}
\label{sec-moon}

We use the same lunar model as in \citet{Gevorgyan2023} and
\citet{GevorgyanPackage2026} to analyze the differential map of its effective
homogeneous rheology.

We first use the full transfer matrix to determine which solid-layer
rheological perturbations are visible in the complete Love-number response
when expressed in effective-rheology coordinates. We then restrict the analysis
to the frequency interval relevant to lunar laser ranging and determine
which perturbations remain visible to the observations. Both analyses are
local to the chosen baseline. We do not perform a new inversion of the lunar
interior.

\subsection{Baseline model}
\label{subsec-moon-baseline}

The baseline consists of a solid inner
core, a liquid outer core, a low-viscosity zone, an upper mantle, and a crust.
We model the low-viscosity zone as a viscosity contrast at the base of the
mantle rather than as the low-rigidity transition layer of
\citet{Matsuyama2016} (Table~\ref{tab-lunar-baseline}).

\begin{table*}
\centering
\caption{Baseline parameters for the five-layer lunar Maxwell model.
Radii, densities, and rigidities follow \citet{Matsuyama2016}. Viscosities are
taken from \citep{Gevorgyan2023,GevorgyanPackage2026}.}
\label{tab-lunar-baseline}
\begin{tabular}{@{}lcccc@{}}
\toprule
Layer & Outer radius & Density & Rigidity & Viscosity\\
 & km & kg m$^{-3}$ & GPa & Pa s\\
\midrule
Solid inner core & 213.52 & 7720.16 & 40 & $10^{21}$\\
Liquid outer core & 325.00 & 6700.00 & $10^{-10}$ & 1\\
Low-viscosity zone & 500.00 & 3800.00 & 60 & $5\times10^{16}$\\
Upper mantle & 1697.15 & 3356.00 & 62.5 & $10^{21}$\\
Crust & 1737.15 & 2735.00 & 15 & $10^{23}$\\
\bottomrule
\end{tabular}
\end{table*}

We vary only the viscosities and rigidities of the solid viscoelastic layers:
\begin{equation}\label{eq-vfull-moon}
    v_{\rm full}
    =
    (\eta_{\rm ic},\mu_{\rm ic},
    \eta_{\rm LVZ},\mu_{\rm LVZ},
    \eta_{\rm man},\mu_{\rm man},
    \eta_{\rm cr},\mu_{\rm cr})^T.
\end{equation}
The liquid outer core, radii, and densities are fixed. All diagnostics below
use relative perturbations as in \eqref{eq-relative-vars}.

The lunar response has $n+1=9$ relaxation poles, hence $n=8$ Voigt
elements plus the spring--dashpot pair $(\alpha,\eta)$. We distinguish
mode times $\tau_i=-1/s_i$ from the retardation times
$\eta_i/\alpha_i$.

\subsection{Full map and resolution}
\label{subsec-moon-full}

\begin{figure}
\centering
\includegraphics[width=\columnwidth]{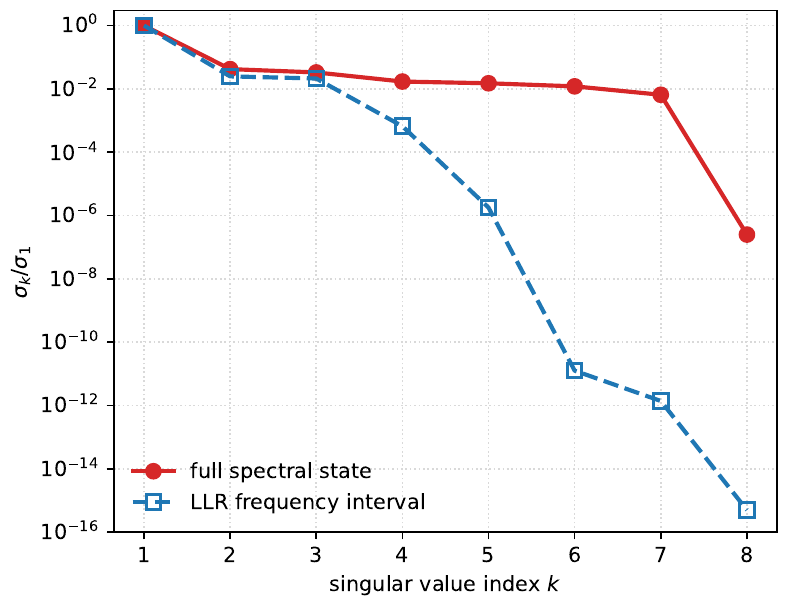}
\caption{Singular-value spectra at the lunar baseline, normalized by the
largest value. Filled circles show the full scaled transfer matrix
$\widehat T_{\rm full}$ \eqref{eq-moon-full-T}. Open squares show the
frequency-interval physical map $\widehat J_{v,\mathcal B}$ discussed in
Section~\ref{subsec-moon-band}. Lines connect ordered singular values. The full map has a clear gap after the seventh singular
value. Fewer directions remain visible in the LLR frequency interval.}
\label{fig-moon-scree}
\end{figure}

For this baseline, we omit the fixed $\gamma$ row and evaluate the scaled full transfer matrix
\begin{equation}\label{eq-moon-full-T}
    \widehat T_{\rm full}
    =
    \diag(w_0)^{-1}D\mathcal I(u_0)J_v\diag(v_0).
\end{equation}
We normalize its singular values by the largest one:
\begin{equation}\label{eq-moon-svd}
\begin{aligned}
    \sigma_k/\sigma_1=\{&1,\;0.042,\;0.033,\;0.017,\\
    &0.015,\;0.012,\;0.0065,\;2.5\times10^{-7}\}.
\end{aligned}
\end{equation}
The first seven singular values remain within about two orders of magnitude.
The eighth drops by more than four orders (Figure~\ref{fig-moon-scree}).
We therefore find eight physical directions, seven response-sensitive and
one nearly invisible.

The truncated resolution matrix,
\begin{equation}\label{eq-moon-res}
    R_{\rm full}
    =
    \widehat T_{{\rm full},\tau}^{+}\widehat T_{\rm full}
    =
    V_7V_7^T,
\end{equation}
is formed by placing the threshold $\tau$ in the gap between
$\sigma_7/\sigma_1$ and $\sigma_8/\sigma_1$. Seven diagonal entries are equal
to one within rounding error. The inner-core-rigidity entry is
$R_{\mu_{\rm ic},\mu_{\rm ic}}\approx4\times10^{-7}$. The discarded right
singular vector is aligned with the inner-core-rigidity axis to better than one
part in $10^6$.

\begin{figure}
\centering
\includegraphics[width=\columnwidth]{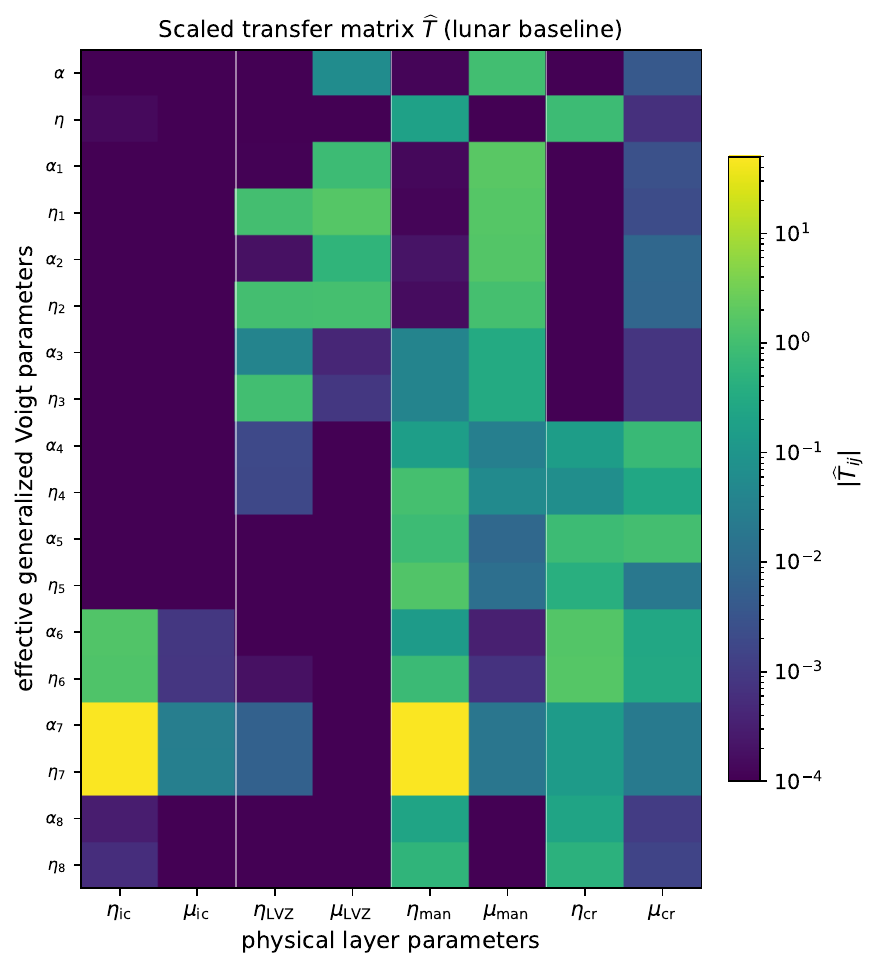}
\caption{Magnitude of the scaled transfer matrix $\widehat T_{\rm full}$ at the
lunar baseline. The color scale is logarithmic, and the vanishing $\gamma$ row
is omitted. Rows
are effective generalized Voigt parameters and columns are the eight physical
rheological parameters. The inner-core rigidity column $\mu_{\rm ic}$ is
uniformly near zero. The remaining columns contain non-negligible entries.}
\label{fig-moon-heatmap}
\end{figure}

This alignment is consistent with the
weak dependence of Love numbers on core elastic structure
\citep{OkuboSaito1983} and with lunar inversions in which degree-two tidal data
do not resolve the inner core \citep{Briaud2023}. In the transfer-matrix
heatmap (Figure~\ref{fig-moon-heatmap}), the $\mu_{\rm ic}$ column is uniformly
near zero. The other solid-layer parameters produce non-negligible patterns
across the effective parameters. The inner-core viscosity
remains visible because it controls a slow relaxation mode, as shown by the
mode-dominance diagnostic in Section~\ref{subsec-moon-dominance}.

The large value $\sigma_1=91$ reflects a near-degenerate pair of Voigt elements
with times near $1.2$--$1.3$~Myr rather than an additional loss of rank. Along
the corresponding singular direction, the inner-core and upper-mantle
viscosities vary together and control the separation between the two relaxation
times. When
$\eta_{\rm ic}$ is increased from $10^{21}$ to $3\times10^{21}$~Pa\,s, the
largest singular value drops to $3.8$. The seven-plus-one split remains
unchanged under this and the other baseline-viscosity changes in
Table~\ref{tab-moon-robustness}. Changing the LVZ or crust viscosity shifts the
corresponding relaxation times but does not change the rank structure of the
full transfer map or the alignment of the near-null direction with
$\mu_{\rm ic}$.

\begin{table*}
\centering
\caption{Robustness of the main full-map diagnostics under selected changes of
the baseline viscosities. In all cases, the spectrum retains seven response-sensitive
directions and one near-null direction aligned with the inner-core rigidity
axis.}
\label{tab-moon-robustness}
\begin{tabular}{@{}lccl@{}}
\toprule
Variant & $\sigma_8/\sigma_1$ & $R_{\mu_{\rm ic},\mu_{\rm ic}}$ & Main spectral change\\
\midrule
Reference model & $2.5\times10^{-7}$ & $3.8\times10^{-7}$ & ---\\
$\eta_{\rm LVZ}\times10$ & $2.2\times10^{-7}$ & $3.8\times10^{-7}$ & LVZ timescales increase by $10$\\
$\eta_{\rm LVZ}/10$ & $2.5\times10^{-7}$ & $3.8\times10^{-7}$ & LVZ timescales decrease by $10$\\
$\eta_{\rm cr}=10^{21}$~Pa\,s & $3.8\times10^{-7}$ & $3.8\times10^{-7}$ & Crust-controlled modes move to mid-spectrum\\
\bottomrule
\end{tabular}
\end{table*}

\subsection{Mode dominance}
\label{subsec-moon-dominance}

Equation~\eqref{eq-dominance} compares the relative influence of the physical
parameters on each relaxation pole. For a given pole $s_i$, the entries $D_{ij}$
are non-negative and sum to one over $j$. For example, $D_{ij}=0.7$ means that
$v_j$ accounts for $70\%$ of the total absolute first-order sensitivity of
$s_i$ among the selected parameters. At the lunar baseline, the modes separate
into groups associated mainly with the low-viscosity zone, the upper mantle,
the crust, and the inner-core viscosity (Table~\ref{tab-moon-dominance}). We
compute the matrix from pole sensitivities before applying the effective
inversion, so the result concerns physical timescales rather than how they
appear in generalized Voigt parameters.

\begin{table}
\centering
\caption{Relaxation times and dominant rheological controls of each relaxation
mode at the lunar baseline, from the mode-dominance matrix \eqref{eq-dominance}.
The entries are relative pole sensitivities, row-normalized, with contributions
below $2\%$ omitted. The mode times $\tau_i=-1/s_i$ are distinct from the
Voigt retardation times $\eta_i/\alpha_i$. Asterisks mark the two modes whose
associated elements are retained by the
frequency-interval model in
Section~\ref{subsec-moon-band}.}
\label{tab-moon-dominance}
\footnotesize
\begin{tabular}{@{}ccp{0.58\columnwidth}@{}}
\toprule
Mode & $\tau_i$ & Dominant parameters (\%)\\
\midrule
$s_1^\ast$ & $12.2$~d & $\eta_{\rm LVZ}\,50$, $\mu_{\rm LVZ}\,39$, $\mu_{\rm man}\,11$\\
$s_2^\ast$ & $18.3$~d & $\eta_{\rm LVZ}\,50$, $\mu_{\rm LVZ}\,26$, $\mu_{\rm man}\,24$\\
$s_3$ & $63$~yr & $\eta_{\rm LVZ}\,99.8$\\
$s_4$ & $1.9\times10^{4}$~yr & $\mu_{\rm cr}\,49$, $\eta_{\rm man}\,46$, $\eta_{\rm cr}\,4$\\
$s_5$ & $2.3\times10^{4}$~yr & $\eta_{\rm man}\,80$, $\mu_{\rm cr}\,16$\\
$s_6$ & $2.6\times10^{5}$~yr & $\eta_{\rm cr}\,54$, $\mu_{\rm cr}\,45$\\
$s_7$ & $1.29\times10^{6}$~yr & $\eta_{\rm man}\,70$, $\eta_{\rm ic}\,29$\\
$s_8$ & $1.32\times10^{6}$~yr & $\eta_{\rm ic}\,71$, $\eta_{\rm man}\,29$\\
$s_9$ & $4.0\times10^{8}$~yr & $\eta_{\rm cr}\,91$, $\eta_{\rm man}\,9$\\
\bottomrule
\end{tabular}
\end{table}

Each row shows which physical parameters most strongly change a pole at first
order. The absence of $\mu_{\rm ic}$ is consistent with its near-zero
resolution in Section~\ref{subsec-moon-full}. The presence of $\eta_{\rm ic}$
in $s_8$ explains why the inner-core viscosity remains visible in the full
response.

\subsection{Frequency-interval reduction}
\label{subsec-moon-band}

The complete pole-residue state represents the full effective rheology. For the
Moon, we restrict the analysis to the frequency interval covered by lunar laser
ranging (LLR) constraints,
$\omega\in[3.3\times10^{-8},2.7\times10^{-6}]~{\rm s}^{-1}$, corresponding to
forcing periods from about one month to six years
\citep{WilliamsBoggs2015,Gevorgyan2023}. At frequencies $\omega_m$ across
the interval, we evaluate
\begin{equation}\label{eq-band-operator}
\begin{aligned}
    \mathcal B(u)&=\bigl(k_2(i\omega_1),\ldots,k_2(i\omega_M)\bigr),\\
    k_2(i\omega)&=k_E+\sum_{i=1}^{n+1}\frac{r_i}{i\omega-s_i},
\end{aligned}
\end{equation}
where $M=50$ frequencies are logarithmically spaced across the LLR frequency
interval. The Jacobians contain separate rows for the real and imaginary parts. The derivative of the sampling operator has rows
\begin{align}\label{eq-band-rows}
    \frac{\pp k_2(i\omega_m)}{\pp k_E}&=1,\\
    \frac{\pp k_2(i\omega_m)}{\pp s_i}
    &=\frac{r_i}{(i\omega_m-s_i)^2},\\
    \frac{\pp k_2(i\omega_m)}{\pp r_i}
    &=\frac{1}{i\omega_m-s_i}.
\end{align}

The physical and effective frequency-interval maps are then
\begin{equation}\label{eq-band-maps}
    J_{v,\mathcal B}=D\mathcal B(u_0)J_v,
    \quad
    J_{w,\mathcal B}=D\mathcal B(u_0)J_w.
\end{equation}
A reduced model contains a selected subset of the parameter pairs
$(\alpha,\eta)$ and $(\alpha_i,\eta_i)$. We retain the corresponding columns of
$J_{w,\mathcal B}$ before forming the pseudoinverse in \eqref{eq-residual}.
We evaluate the projection residual $\varepsilon_{\mathcal B}$ from
\eqref{eq-residual} with $\mathcal G=\mathcal B$. The number of required
pairs depends strongly on the response being matched. Against the full spectral state,
omitting any pair changes the sensitivity at the percent level. In the LLR frequency
interval, the slow modes contribute negligibly to the first-order
sensitivity. Keeping only the global
spring--dashpot pair and the two elements associated with the starred modes
$s_1^\ast$ and $s_2^\ast$ in Table~\ref{tab-moon-dominance} gives
\begin{equation}
    \varepsilon_{\mathcal B}=2\times10^{-7}.
\end{equation}

The real-part Jacobian is $19.1$ times larger than the imaginary-part Jacobian
in the Frobenius norm. The componentwise residuals
\eqref{eq-component-residuals} are
\begin{equation}
    \varepsilon_{\mathcal B}^{\operatorname{Re}}
    =1.7\times10^{-7},
    \qquad
    \varepsilon_{\mathcal B}^{\operatorname{Im}}
    =1.6\times10^{-6}.
\end{equation}
The dissipative first-order sensitivity is reproduced
to about two parts in $10^6$ over the interval.

\begin{figure}
\centering
\includegraphics[width=\columnwidth]{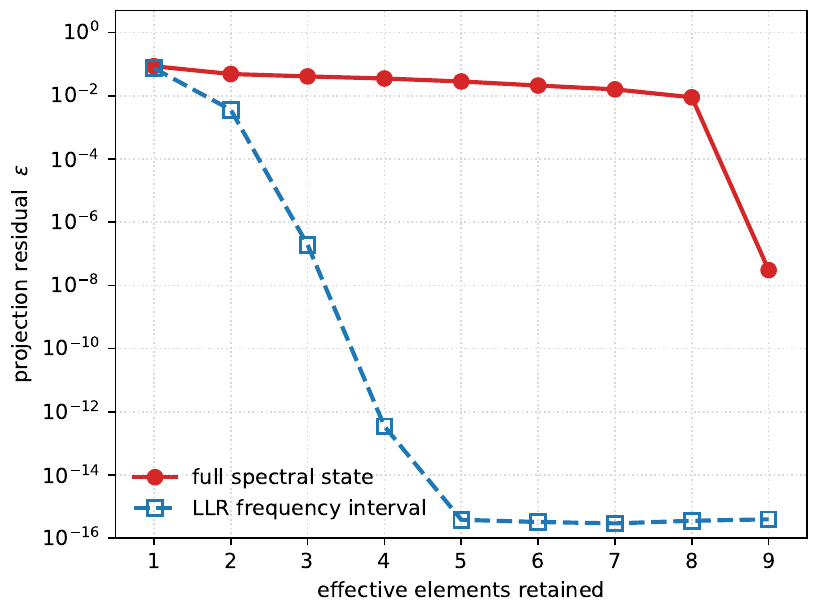}
\caption{Projection residual \eqref{eq-residual} for nested reduced effective
models obtained by successively discarding the least influential parameter
pairs. The connected points give the resulting sequence of reduced models.
The full spectral response retains sensitivity to all pairs. Over the
LLR frequency interval, the pair $(\alpha,\eta)$ and the two elements
associated with the starred modes listed in Table~\ref{tab-moon-dominance}
reproduce the response.}
\label{fig-moon-duality}
\end{figure}

The pair $(\alpha,\eta)$ must be retained because $\alpha$ carries the elastic
part of the response. Freezing this pair gives a frequency-interval residual of order
$10^{-1}$. The reduced frequency-interval model consists of this pair and the two
elements associated with the 12.2 and 18.3 d Love-number modes
(Figure~\ref{fig-moon-duality}).

The omitted modes describe slower relaxation processes, including the
viscosity directions of the inner core and crust identified in
Table~\ref{tab-moon-dominance}. Their contributions to the Jacobian over the
chosen interval are negligible. The singular spectrum of
$\widehat J_{v,\mathcal B}$ in Figure~\ref{fig-moon-scree} contains
fewer visible physical directions than the full effective map.

\begin{figure}
\centering
\includegraphics[width=\columnwidth]{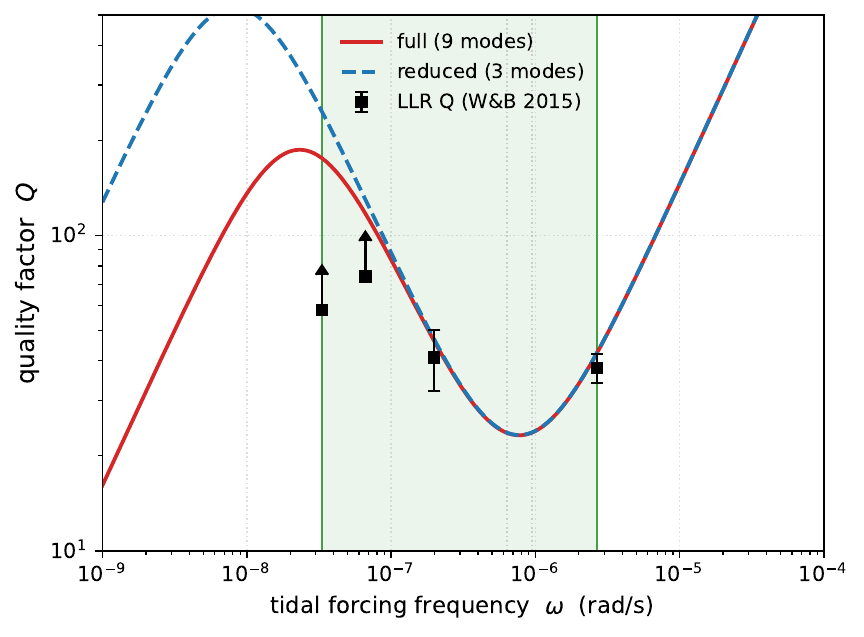}
\caption{Lunar tidal quality factor $Q(\omega)$ for the full nine-mode
effective model and for the reduced frequency-interval model retaining the pair $(\alpha,\eta)$ and
the two elements whose modes lie within the interval. The fit uses the complex Love
number $k_2(i\omega)$. The reduced model agrees with the full
response through the LLR frequency interval and departs below its lower bound, where
the omitted slow modes begin to contribute. Points are the lunar laser
ranging determinations of \citet{WilliamsBoggs2015}, compiled by
\citet{Gevorgyan2023}.}
\label{fig-moon-response}
\end{figure}

We plot the resulting response as $Q(\omega)$ in
Figure~\ref{fig-moon-response}.
The reduced model follows the full model through the
LLR frequency interval and deviates below its lower bound, where the omitted slow
modes begin to affect the response. We fit the complex Love number
$k_2(i\omega)$ in \eqref{eq-band-operator}.

\subsection{Finite-perturbation check}
\label{subsec-moon-finite}

We check the tangent prediction against finite perturbations of the
physical parameters. For each coordinate direction $e_j$, we perturb the
baseline as
\begin{equation}
    v=v_0+\delta v_{0,j} e_j,
\end{equation}
rerun the full pointwise reduction, and compare the recomputed effective vector
with the linear prediction
\begin{equation}
    w_0+T\Delta v.
\end{equation}

To quantify the error, we compare relative changes in the complete effective
parameter vector. For physical coordinate $j$, define
\begin{equation}
\begin{aligned}
    \delta w_j^{\rm num}
    &=\diag(w_0)^{-1}
      \left[w(v_0+\delta v_{0,j}e_j)-w_0\right],\\
       \delta w_j^{\rm lin}
    &=\delta\widehat T e_j,
\end{aligned}
\end{equation}
We measure the relative error as
\begin{equation}\label{eq-finite-error}
    E_j(\delta)
    =\frac{\|\delta w_j^{\rm num}-\delta w_j^{\rm lin}\|_2}
    {\|\delta w_j^{\rm num}\|_2}.
\end{equation}
Because a perturbation can reorder the Voigt elements, we pair each recomputed
element with the baseline element having the nearest retardation time
$\eta_i/\alpha_i$ on a logarithmic scale. The errors quoted below are values of
$E_j$. The maximum is taken over the eight physical parameters.

The prediction matches the recomputed effective parameters to about $2\%$
  or better for $\delta=10^{-3}$. At $\delta=10^{-2}$, six of the eight
  coordinate directions remain accurate to about $1\%$, but the inner-core and
  upper-mantle viscosity directions degrade to roughly $20\%$, because they
  control the nearly degenerate $1.2$--$1.3$~Myr element pair discussed in
  Section~\ref{subsec-moon-full}. The tangent approximation remains within a
  few percent for $\delta\lesssim10^{-3}$. For larger perturbations, errors first
  increase in these two viscosity directions, and the ordering of the two
  elements can change near $\delta=10^{-1}$.

The fixed-geometry constraint is preserved throughout the finite-perturbation
tests: the recomputed $k_f$ and $\gamma$ remain constant to better than
$2\times10^{-6}$. A $10\%$ perturbation of the inner-core rigidity changes the effective vector about
thirty times less than the same perturbation applied to any other physical
parameter.

\section{Discussion and conclusions}
\label{sec-discussion}

At the level of the frequency-dependent Love number, a finite multilayer
Maxwell body can be represented by a homogeneous body with generalized rheology
\citep{Gevorgyan2023}. Here, we differentiate the mapping about a chosen
layered baseline. The resulting local map quantifies how perturbations of the
layer rheologies change the effective homogeneous parameters and which
perturbations have little influence on the selected response.

We can use the transfer matrix for more than interpreting the effective
parameters. Once an effective model has been constructed at $v_0$, we estimate
its response to prescribed small changes in the layer rheologies from
$\Delta w\approx T\Delta v$. This local update avoids repeating the layered
Love-number calculation and the algebraic inversion at every step of a
downstream orbital or rotational calculation. The update
is valid only within the local range established by the finite-perturbation
tests, and the baseline map must be recomputed when the evolving model leaves
that range. If the uncertainties and correlations of the physical parameters
are described by the covariance matrix $\Sigma_v$, then the corresponding
effective-parameter covariance is, to first order,
\begin{equation}\label{eq-effective-covariance}
    \Sigma_w\approx T\Sigma_vT^{\mathsf T}.
\end{equation}
We use the singular values and projection residuals to select a compact
effective parameter set that retains the physical sensitivity over a chosen
frequency interval.

Standard Love-number sensitivities directly describe how physical parameters
perturb $k_2$. We instead use the transfer matrix to express that local response
change in the parameters of an effective rheology. A homogeneous time-domain
solver can use the updated rheology and its uncertainty directly. The
transfer matrix is not, by itself, an inversion of observations for interior
parameters. Such an inference additionally requires an observational operator,
data covariance, prior information, and regularization.

We distinguish quantities determined by the Love-number response from
quantities that depend on the effective parameterization. The elastic limit,
fluid limit, relaxation poles, and residues depend only on the Love-number
response. If we change the effective parameterization, the numerical entries
of the transfer matrix change. For example, generalized Voigt, Burgers, and
finite Andrade models can describe the same local response using different
parameter sets \citep{Bland2016,Gevorgyan2021,Gevorgyan2023}. The rank,
physical near-null directions, and projection residuals therefore depend less
on the chosen parameterization than the individual matrix entries.

For the complete effective representation, $D\mathcal I(u_0)$ is nonsingular on
the rheological tangent space defined by fixed geometry and density. Consequently,
\begin{equation}
    \ker T=\ker J_v,
\end{equation}
and the homogeneous change of coordinates introduces no additional exact null
directions. The near-null inner-core rigidity direction found in the full map
reflects the weak sensitivity of the Love-number response itself. Additional
response information is lost when we analyze only part of the response or omit
effective elements. Truncating small singular directions affects only the
diagnostic resolution adopted for interpretation.

In the lunar model, the full transfer map resolves seven of the eight solid-layer
rheological directions and isolates the inner-core rigidity as the single
near-null direction. This result agrees, in effective-rheology coordinates,
with the weak sensitivity of degree-two Love numbers to the elastic structure
of the inner core \citep{OkuboSaito1983,Briaud2023}. We obtain the same seven-plus-one split
after the selected changes to the baseline viscosities.

We use the mode-dominance diagnostic to determine which physical rheological
parameters most strongly move each pole at first order, without assigning each
effective mode to a single geological layer. In the lunar model, the
low-viscosity zone, mantle, crust, and \mbox{inner-core viscosity} affect distinct
relaxation times. Over the frequency interval constrained by lunar laser
ranging, the first-order response is reproduced by the pair $(\alpha,\eta)$ and
the two Voigt elements associated with the modes within the interval.
Contributions from the slower modes are negligible over this interval.

The equivalence considered here is an equivalence of the external linear
response, not of the internal deformation fields. Matching the complex Love
number reproduces the surface response and the total degree-two tidal
dissipation for the same forcing, but it does not reproduce the radial
distribution of stress and strain. Consequently, the layer-wise or local
dissipation cannot be recovered uniquely from the effective homogeneous
parameters alone. Different layered interiors can produce nearly identical
Love numbers but partition the dissipated energy differently among their
layers.

This distinction between global and local dissipation is well established.
\citet{Tobie2005} compute the global dissipated power from
$\operatorname{Im}k_2$ and determine the radial distribution of dissipation
using sensitivity functions derived from the radial deformation functions.
\citet{Beuthe2013} represents spatial heating in a spherically stratified body
using radial weight functions that depend on the internal structure and
angular patterns determined by the tidal forcing. Integrating the corresponding
local dissipation density over each layer gives the layer fractions used below.

In a coupled orbital--thermal model, we could supplement the effective rheology
with dissipation-partition information calculated from the layered baseline
using such radial sensitivity or weight functions. If
$f_\ell(\omega;v)$ is the fraction of the total dissipation
deposited in layer $\ell$, the simplest local approximation is
\begin{equation}\label{eq-layer-dissipation-fraction}
    \dot E_\ell(\omega;v)
    \approx
    f_\ell(\omega;v_0)\,
    \dot E_{\rm tot}^{\rm eff}(\omega;v),
\end{equation}
where the effective model supplies the total dissipation. For a first-order
update, we may use
\begin{equation}\label{eq-layer-fraction-linearization}
    f_\ell(\omega;v_0+\Delta v)
    \approx
    f_\ell(\omega;v_0)
    +D_vf_\ell(\omega;v_0)\Delta v.
\end{equation}
The heating prescription remains local to the selected baseline and requires
additional layered-model calculations. The partition functions
are not contained in the Love number or in the transfer matrix developed here.

The retained poles must be simple and remain separated under perturbation. As
poles approach degeneracy, individual
pole derivatives become ill-conditioned, and one should work with
invariant subspaces rather than individual modes \citep{Nelson1976}. At the
lunar baseline, the nearly degenerate $1.2$--$1.3$~Myr element pair limits the
range of the tangent approximation. We assume a rational response. This is
exact for finite multilayer Maxwell bodies \citep{Sabadini2016,Gevorgyan2023}.
Rheologies with continuous relaxation spectra, such as Andrade or
Sundberg--Cooper models, give non-rational Love numbers. The finite-dimensional
differential method developed here can be applied after representing the
response by a finite generalized Voigt model over the frequency interval
of interest \citep{Gevorgyan2020,GevorgyanRagazzo2026}.

An observation-limited extension would combine the Jacobians derived here
with an observation operator to determine which layered and homogeneous
response directions can be distinguished from finite data, such as
measurements of $k_2$, $Q$, $k_2/Q$, moment of inertia, or libration amplitude.

Effective homogeneous parameters are therefore meaningful only relative
to a baseline, a parameterization, and the response quantity being considered.

\section*{Acknowledgements}

This work was partially supported by King Abdullah University of Science and Technology (KAUST), Saudi Arabia.

\section*{Data Availability}

No new observational data were generated for this study. The code used to
compute the spectral sensitivities, transfer matrices, diagnostics, and figures
is available at \url{https://github.com/ygevorgyan/local-differential-map} and
archived on Zenodo with a citable DOI \citep{GevorgyanLocalDifferentialMapCode2026}.
The homogeneous-rheology construction used as the pointwise reduction workflow
is archived separately \citep{GevorgyanHomogeneousModelCode2026}.

\bibliographystyle{mnras}
\bibliography{mybibliography}

\bsp
\label{lastpage}
\end{document}